\documentclass{moriond}

\def\be{\begin{equation}}
\def\ee{\end{equation}}
\def\bea{\begin{eqnarray}}
\def\eea{\end{eqnarray}}

\newcommand{\vev}[1]{\langle {#1} \rangle}

\newcommand{\eq}[1]{Eq.~(\ref{#1})}

\newcommand{\ord}[1]{\mathcal{O}{(#1)}}
\newcommand{\beq}{\begin{equation}}
\newcommand{\eeq}{\end{equation}}
\newcommand{\eps}{\varepsilon}

\usepackage{graphicx}
\usepackage{amsmath}

\newcommand{\Photo}{\includegraphics[height=35mm]{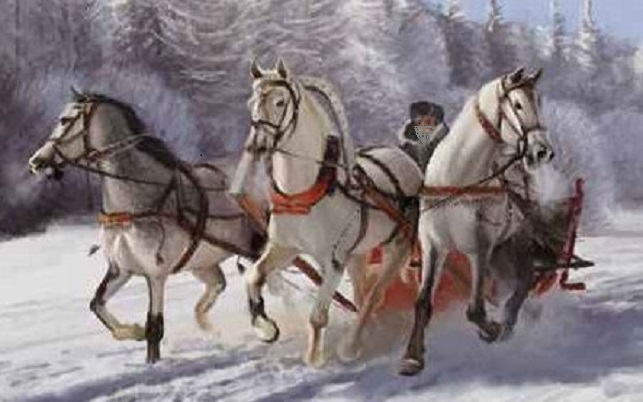}}

\begin{document}
\vspace*{4cm}
\title{MULTI-TEV SIGNALS OF HIGGS TROIKA MODEL}

\author{HOOMAN DAVOUDIASL}

\address{High Energy Theory Group, Physics Department, Brookhaven National Laboratory,\\
Upton, New York 11973, USA}

\maketitle\abstracts{
We consider a model of baryogenesis that requires extending the Standard Model by two additional multi-TeV Higgs doublets that do not break electroweak symmetry.  Adopting the ``Spontaneous Flavor Violation" framework, we can arrange for the heavy Higgs states to have significant couplings to light quarks.  This allows for the heavy scalars to be resonantly produced at a future 100 TeV $pp$ collider and discovered in di-jet and top-pair final states up to masses of $\ord{10~{\rm TeV}}$.  The same mass range can also lead to signals in flavor experiments.  Together, these measurements can play a complementary role in probing the physics involved in this {\it Higgs Troika} baryogenesis scenario.      
}

\section{Introduction}

The following is a summary of the talk given by the author at the virtual Rencontres de Moriond QCD \& High Enrgy Interaction, March 27 - April 3, 2021.  The material is largely based on the results of work performed in collaboration with Ian M. Lewis and Mathew Sullivan, in Refs.~\cite{Davoudiasl:2019lcg,Davoudiasl:2021syn}, where more extensive lists of references relevant to our topic can be found.  The plots presented in this summary are from Ref.~\cite{Davoudiasl:2021syn}

There are a number of open questions in particle physics and cosmology, such as the nature of dark matter, origin of neutrino masses, and the mechanism for generating the baryon asymmetry of the Universe (BAU), that motivate extending the Standard Model (SM).  However, of these problems, the source of the BAU could plausibly have a connection to the visible sector.

Here, we present a model for baryogenesis that assumes two heavy extra Higgs doublets, for a total of three $H_a$, a {\it Higgs Troika}; $a=1,2,3$.  We will identify $H_1$ as that corresponding to observed SM-like Higgs boson.  The new states are in the multi-TeV mass regime.  The model is akin to leptogenesis \cite{Fukugita:1986hr}, in that a $B-L$ asymmetry -- with $B$ and $L$ baryon and lepton numbers, respectively -- is generated by the heavy Higgs decays and processed by the SM electroweak sphalerons into a baryon asymmetry.  The heavy Higgs states decay out-of-equlibrium and the Universe is reheated to $T > 100$~GeV. As in leptogenesis, the electroweak phase transition does not need to be first order.  The couplings of the new Higgs states to SM fermions generally involve new sources of CP violation.  Hence, all Skaharov conditions \cite{Sakharov:1967dj} for successful brayogenesis can be satisfied in the Troika model.  This model can be viewed as a minimal realization of similar proposals in  Refs.~\cite{Dick:1999je,Murayama:2002je,Davoudiasl:2011aa}. More general discussions of three Higgs doublet models can be found, for example, in Refs.~\cite{Cheng:1987rs,Grossman:1994jb,Cree:2011uy,Ivanov:2011ae,Keus:2013hya,Bento:2017eti,Alves:2020brq,Logan:2020mdz}.

\begin{figure}
		\centerline{\includegraphics[width=0.6\linewidth]{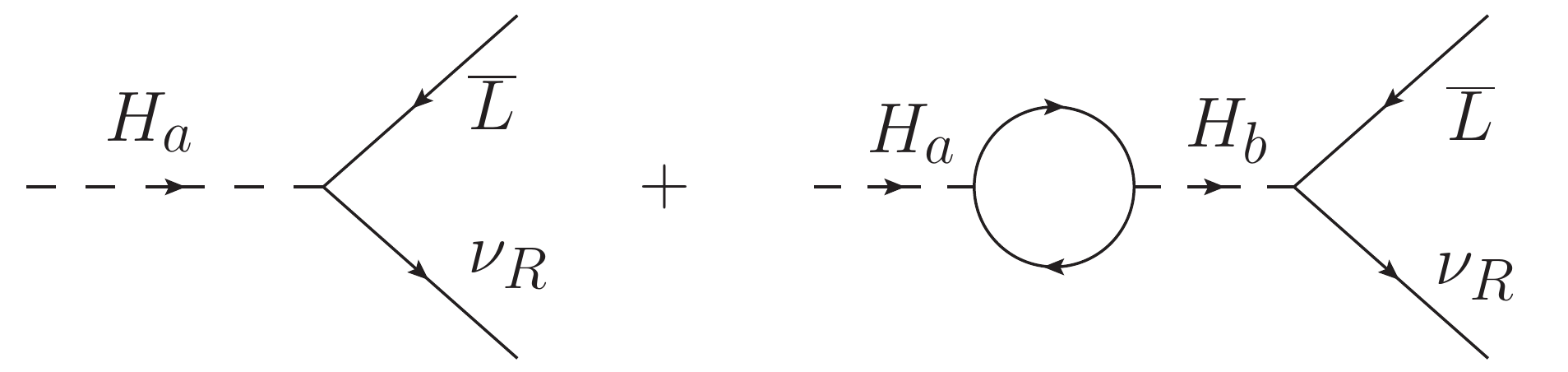}}
\caption[]{Heavy Higgs decay diagrams at tree and the 1-loop level that generate $B-L$ asymmetry.  For degenerate $H_{2,3}$ masses, the other 1-loop processes are not enhanced.}
\label{fig:Hdecay}
\end{figure}

In the above setup, it is assumed that only $H_1$ condenses and breaks electroweak symmetry with a vacuum expectation value $\vev{H_1}=v/\sqrt{2}$; $v=246$~GeV.  The relevant interactions are given by
\beq
\lambda^a_u \tilde H_a^* \bar Q \,u + \lambda^a_d H_a^* \bar Q\, d + 
\lambda^a_\nu \tilde H_a^* \bar L \,\nu_R + 
\lambda^a_\ell H_a^* \bar L \,\ell\,,
\label{Higgs-couplings}
\eeq
where $Q$, $L$, $u$, $d$, $\nu_R$, and $\ell$ are SM quark and lepton doublets, up- and down-type quark singlets, right-handed neutrinos, and charged lepton singlets, respectively; family indices are suppressed.  The asymmetry is generated by the interference of the tree and 1-loop diagrams in Fig.\ref{fig:Hdecay}.  We assume that $H_{2,3}$ masses are somewhat degenerate, at the 5-10\% level and hence the 1-loop ``bubble" diagram is dominant. 

We will assume that the SM neutrinos are Dirac fermions with $\lambda^1_\nu \sim 10^{-12}$.  The asymmetry generated by the $H_a$ decay is defined as
\beq
\eps \equiv \frac{\Gamma(H_a\to \bar L \nu_R) - \Gamma(H_a^*\to \bar \nu_R L)}{2 \Gamma(H_a)},
\label{eps-def}
\eeq  
where $\Gamma (H_a^{(*)}\to X^{(*)})$ denotes the partial with for $H_a^{(*)}\to X^{(*)}$ and $\Gamma(H_a)$ is the total width of $H_a$.  Since we want to generate $\eps \gg \lambda^1_\nu$, the SM Higgs cannot be involved via processes illustrated in Fig.\ref{fig:Hdecay} and two additional Higgs doublets are needed.  Henceforth, we will set $a=2,3$, unless otherwise specified.  We assume that the decay of a heavy modulus $\Phi$ of mass $m_\Phi$ in the early Universe resulted in the production of a population of $H_3$ and $H_3^*$ which then decayed and resulted in a reheat temperature above $T_*\sim 100$~GeV.  One can show \cite{Davoudiasl:2019lcg} that approximately  
\beq
\eps > 3.3 \times 10^{-8} \left(\frac{m_\Phi}{\text{20~TeV}}\right)
\label{eps-mPhi}
\eeq
is required.  After the asymmetry is generated, one needs to make sure it will not be washed out.  For a reheat temperature $\sim T_*$ and one dominant flavor of quarks and leptons contributing to washout, assuming identical $H_a$ couplings, it is found \cite{Davoudiasl:2021syn}
\beq
\lambda_\nu^a \,\lambda_f^a  < 2.1 \times 10^{-4} \left(\frac{m_a}{10~\text{TeV}}\right)^2\,.
\label{Ha-washout}
\eeq

\section{Flavor Model and Constraints}

We will assume that the largest couplings of $H_a$ are to quarks.  In order to have a consistent flavor model for significant coupling of $H_a$ to light quarks, $\sim 0.1$ or larger, we will adapt the the ``Spontaneous Flavor Violation (SFV)" framework of Refs.\cite{Egana-Ugrinovic:2018znw,Egana-Ugrinovic:2019dqu} and focus on the ``up-type" scenario:
\begin{equation}
\label{eq:flavorscheme}
\begin{aligned}
\lambda^{2,3}_u &= \xi \lambda^1_u 
\\
\lambda^{2,3}_d &= \mathrm{diag}( \kappa_{d},\, \kappa_{s},\, \kappa_{b} ) 
\\
\lambda^{2,3}_\ell &= \xi^\ell \lambda^1_\ell 
\\
\lambda^{2,3}_\nu &= \mathrm{diag}( \kappa_{\nu_1},\, \kappa_{\nu_2},\, \kappa_{\nu_3} ) 
\end{aligned}
\end{equation}  
with $H_a$ up-type quark couplings proportional to SM Yukawa couplings, parameterized by constant $\xi$, and down-type quark couplings diagonal, but arbitrary.  The washout condition in \eq{Ha-washout} can be satisfied for $\kappa_{d} \geq 0.01$ and $\kappa_s=\kappa_b=\xi=0$, $\lambda_\nu^{2,3} > 10^{-4}$, and heavy Higgs masses in the range 2-30~TeV.  We then estimate \cite{Davoudiasl:2021syn} the contribution to the electron electric dipole moment (EDM) $d_e$ will always be a factor of 20 or more below the current bound $d_e < 1.1 \times 10^{-28}$~$e$ cm \cite{Andreev:2018ayy}.  Hence, an order of magnitude or more improvement in the bound on $d_e$ can begin to test the model.  

\begin{figure*}[t]
	\includegraphics[width=0.5\columnwidth]{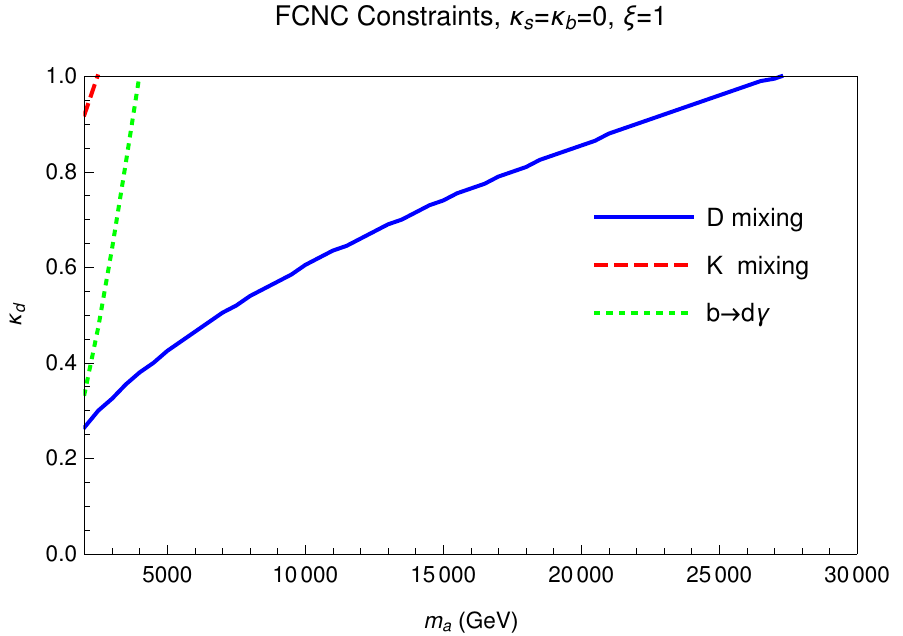}
	\includegraphics[width=0.5\columnwidth]{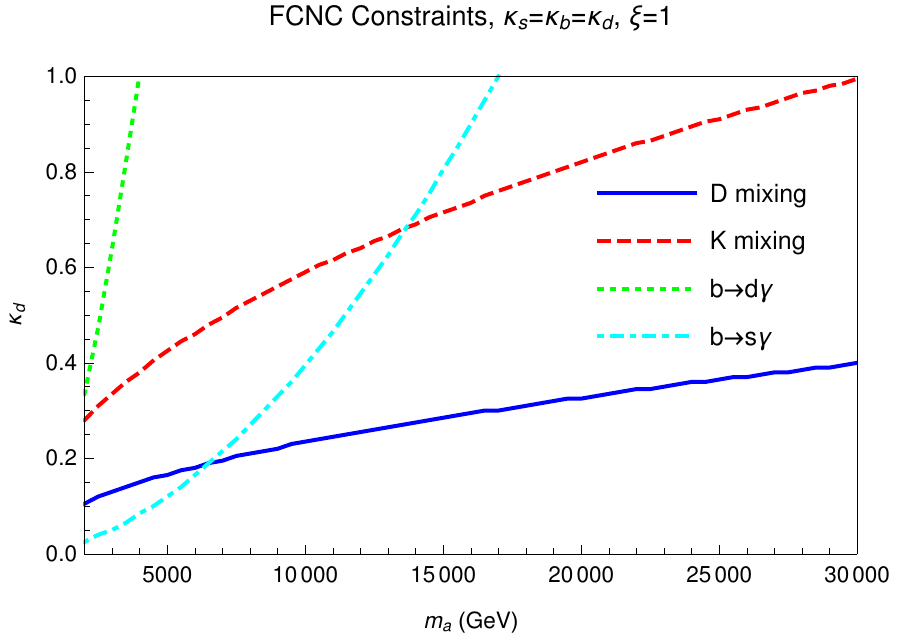}
	\includegraphics[width=0.5\columnwidth]{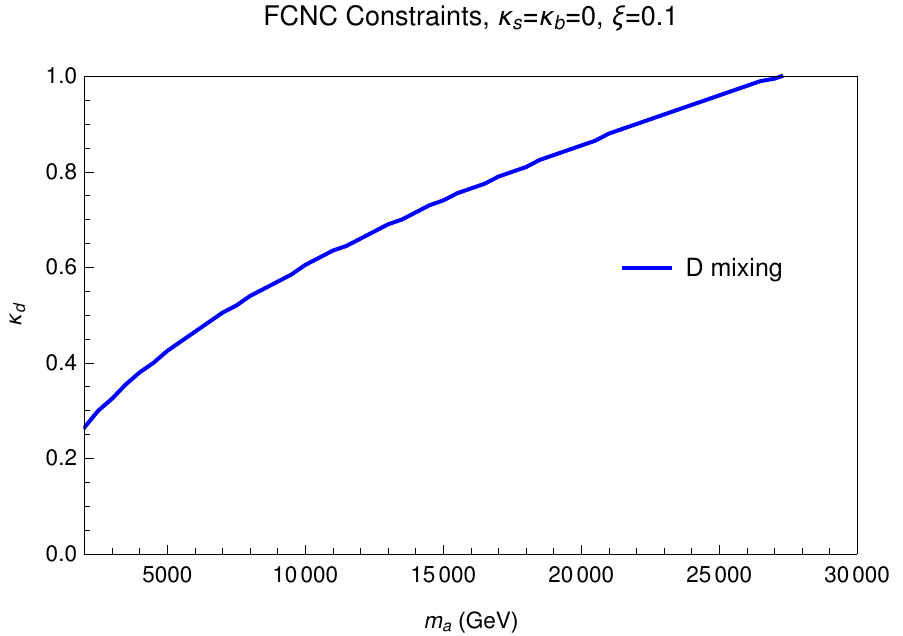}
	\includegraphics[width=0.5\columnwidth]{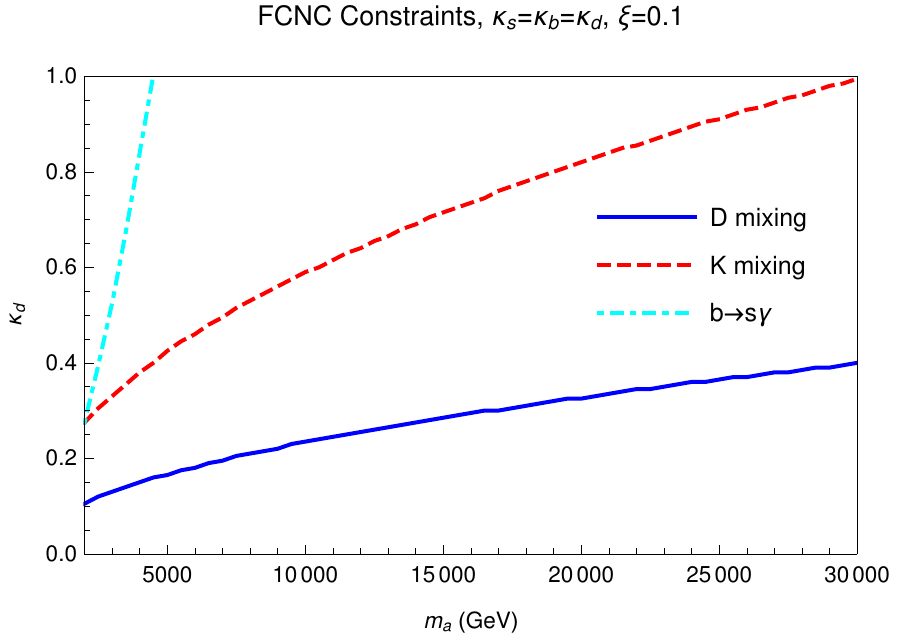}
	\caption{Bounds on the up-type SFV parameters, from flavor data.}
	\label{fig:flavorbounds}
\end{figure*}

Flavor constraints on our model parameters, for different assumptions regarding the couplings, are presented in Fig.\ref{fig:flavorbounds}; for more details see Ref.~\cite{Davoudiasl:2021syn}.  In deriving the bounds, we employ the formalism of Ref.~\cite{Egana-Ugrinovic:2019dqu}  These bounds assume the alignment (no-mixing) limit for the Higgs sector and identical masses and couplings for $H_{2,3}$.  The equality of masses is consistent with our assumption of mild degeneracy of Higgs states and the dominance of the loop diagram in Fig.\ref{fig:Hdecay}.  

\section{Collider Phenomenology}

\begin{figure*}
	\includegraphics[width=0.5\textwidth]{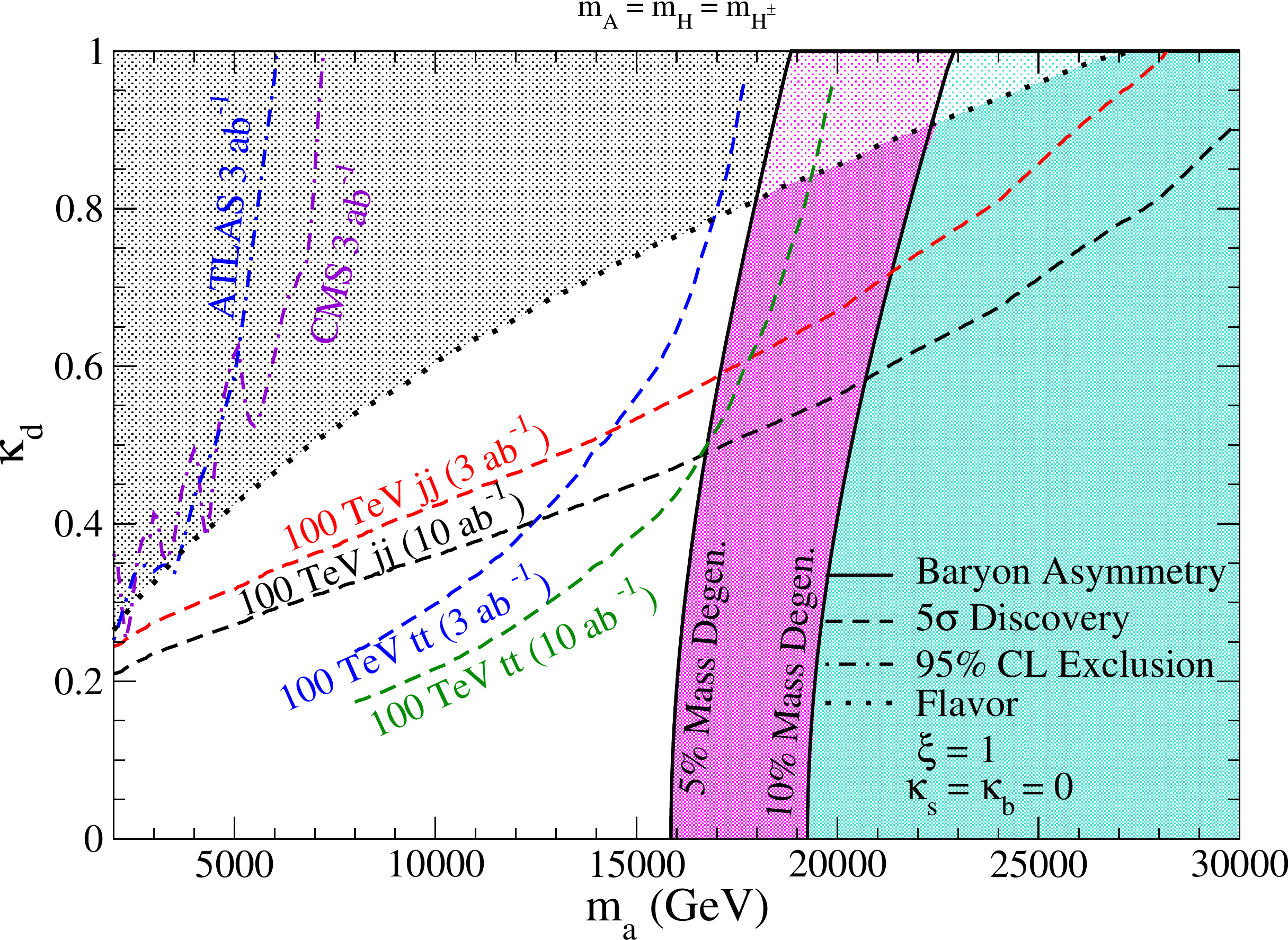}
	\includegraphics[width=0.5\textwidth]{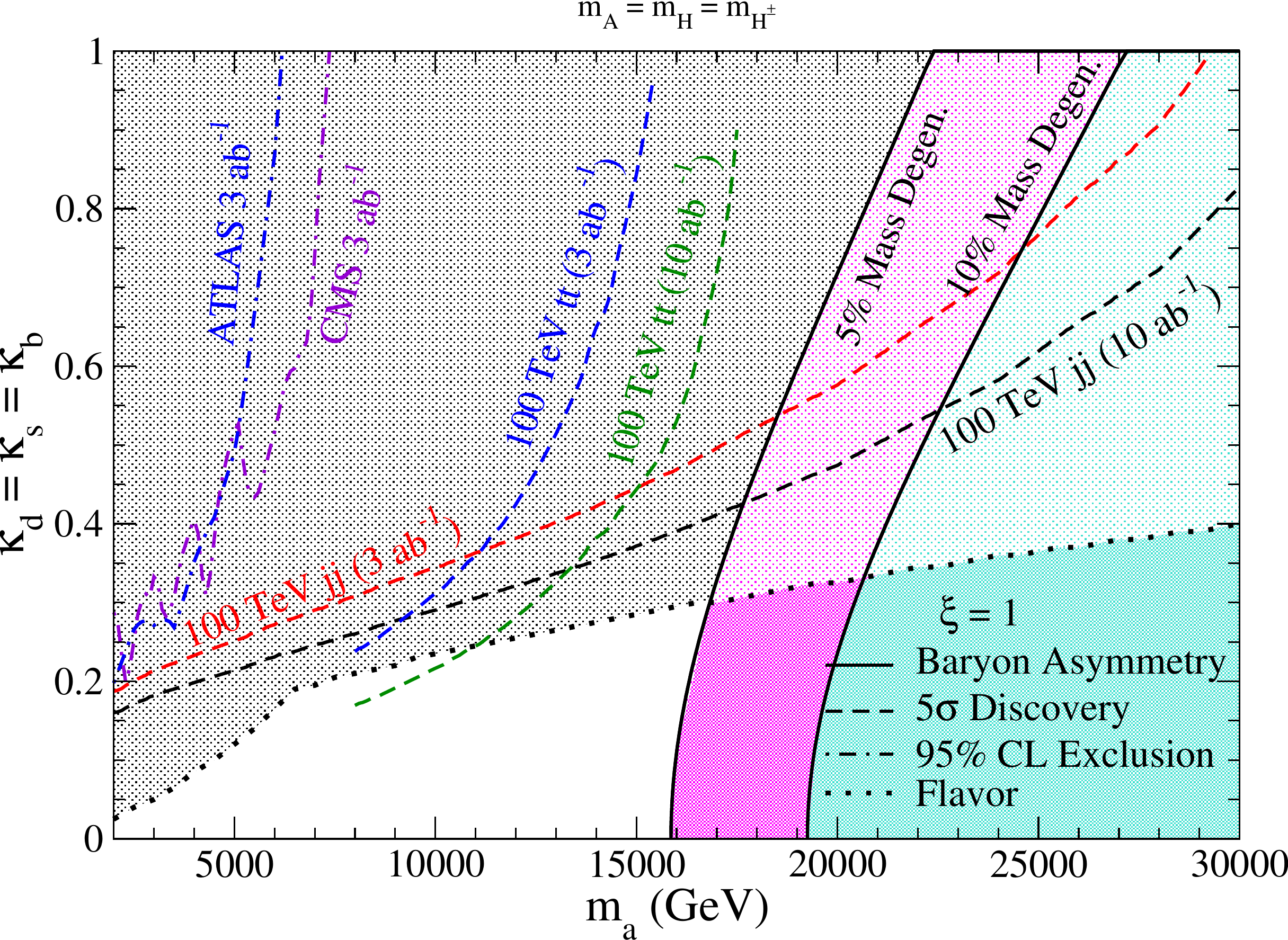}
	\includegraphics[width=0.5\textwidth]{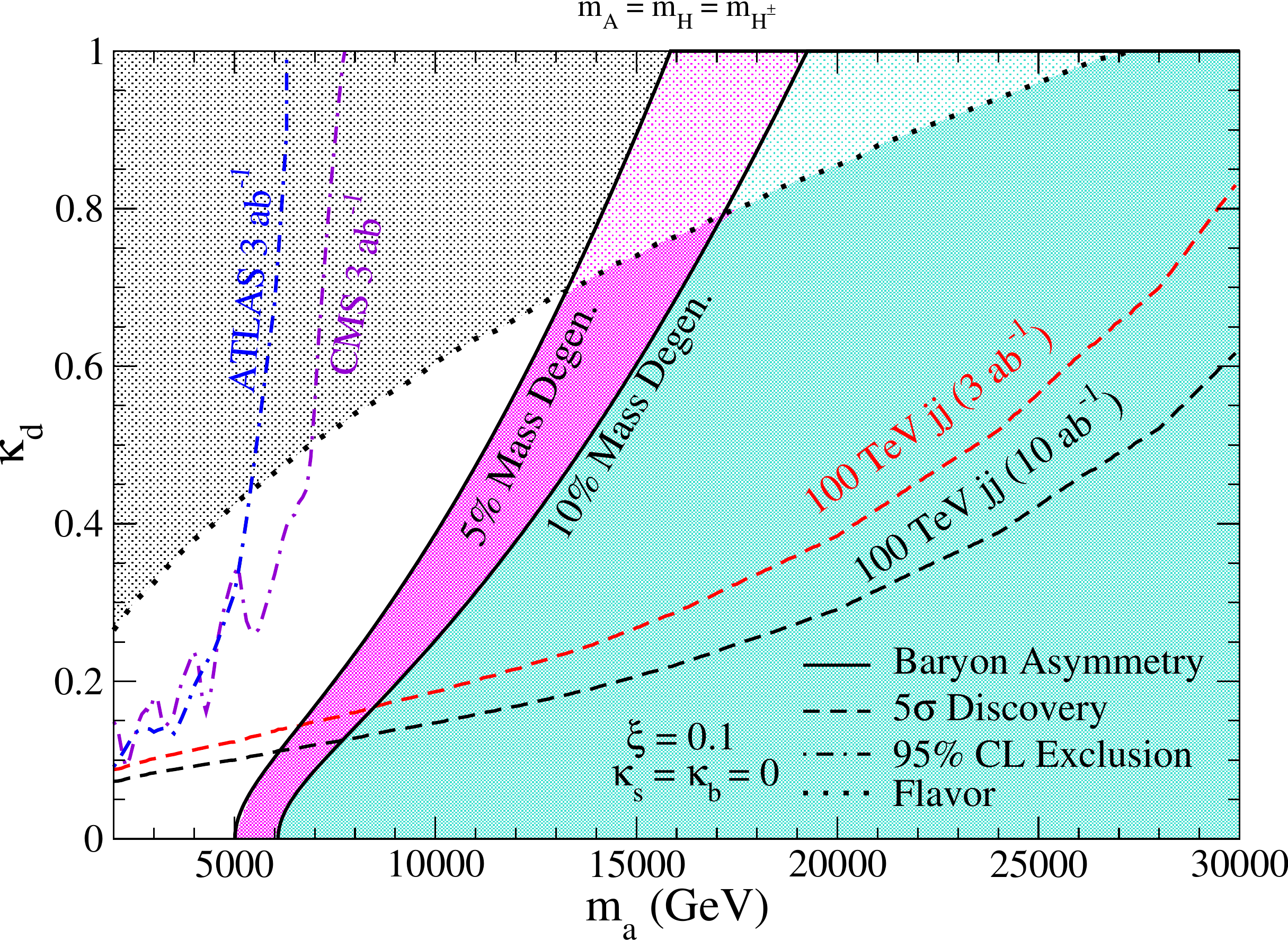}
	\includegraphics[width=0.5\textwidth]{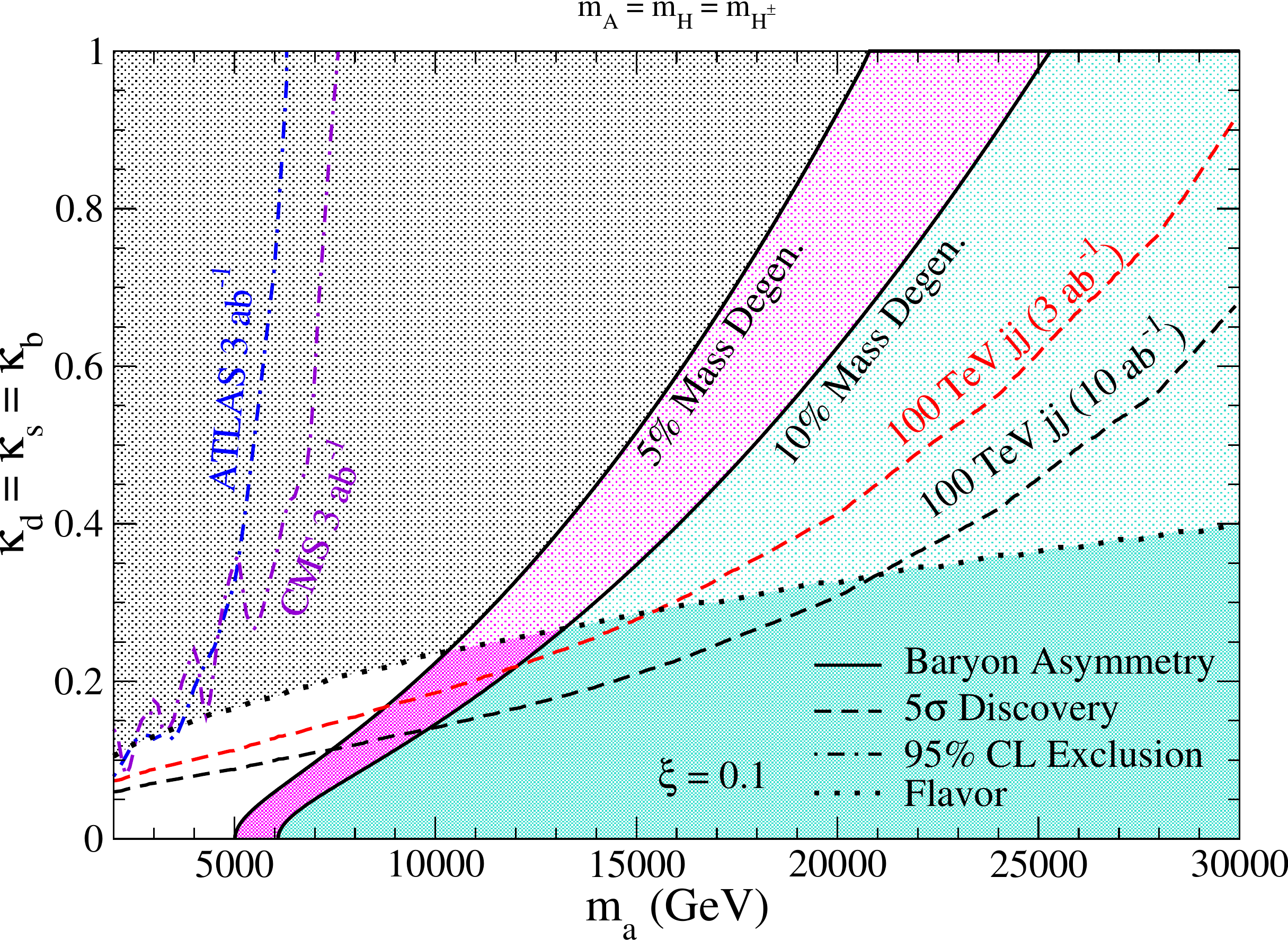}
	\includegraphics[width=0.5\textwidth]{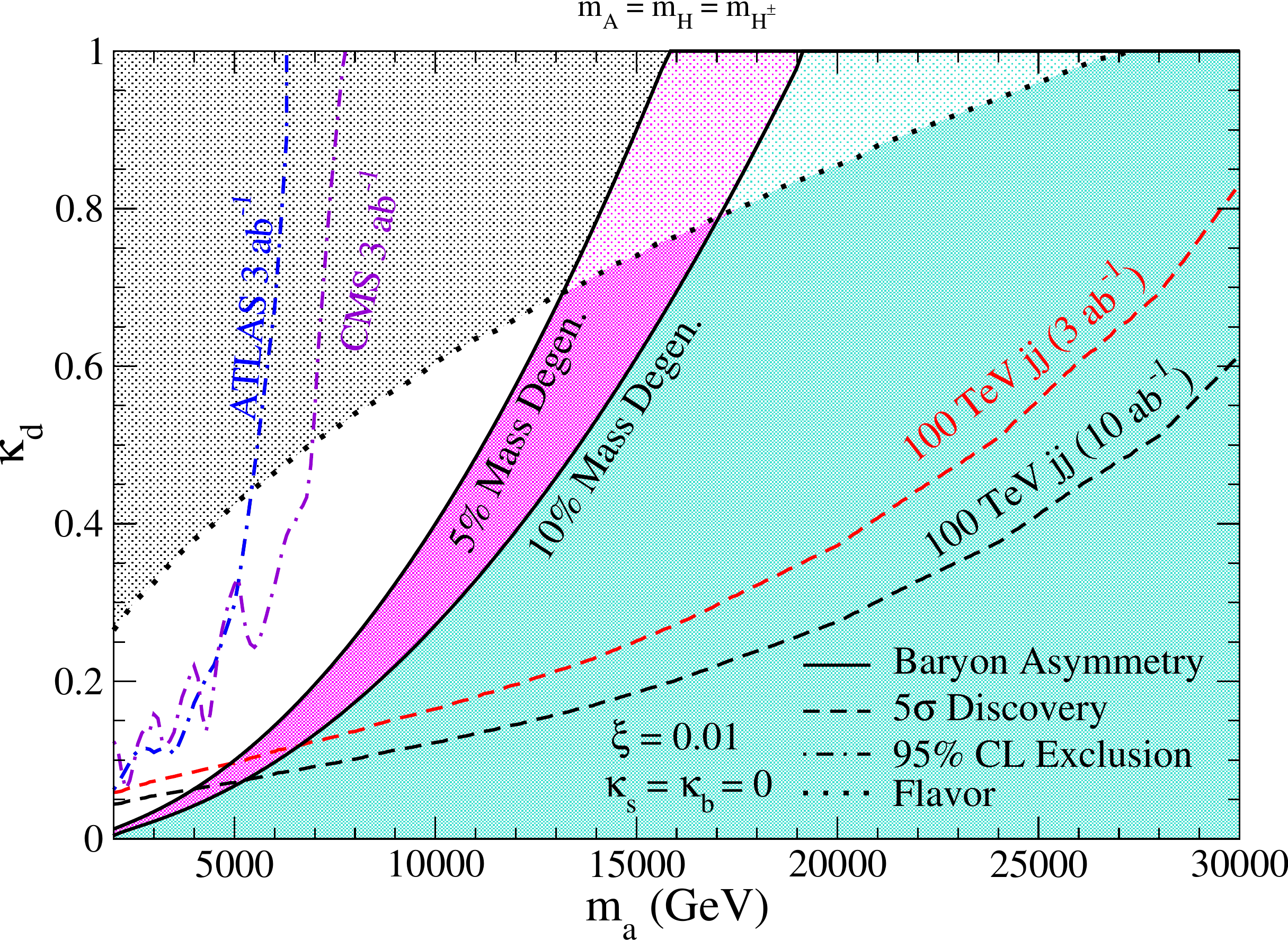}
	\includegraphics[width=0.5\textwidth]{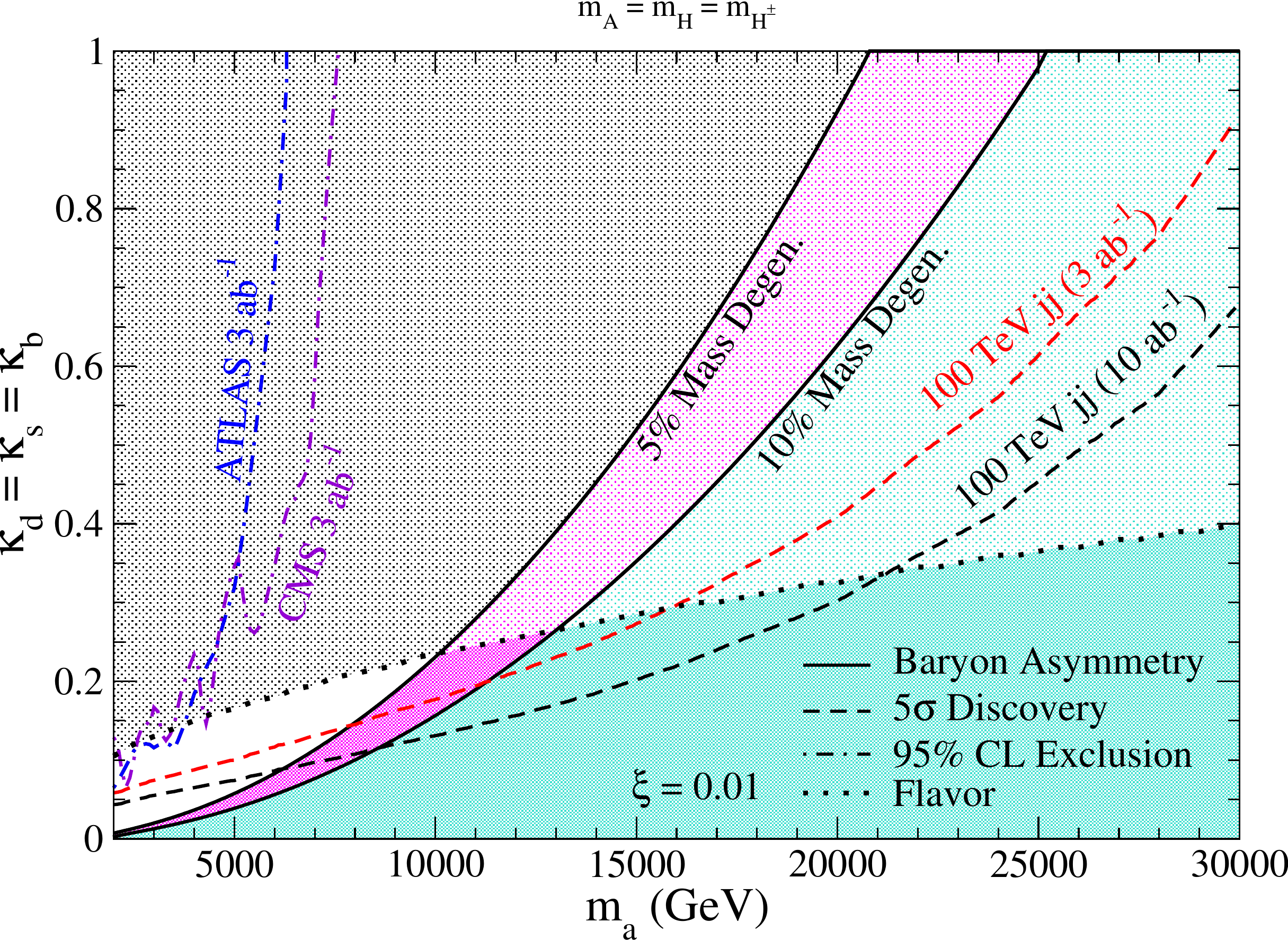}
	\caption{Regions above dashed lines can be discovered (5$\sigma$) in the dijet channel at a 100 TeV $pp$ collider with 3 ab$^{-1}$ (red) and 10 ab$^{-1}$ (black), and the $t\bar{t}$ channel with 3 ab$^{-1}$ (blue) and 10 ab$^{-1}$ (green).  Parameter space above dot-dashed lines can be excluded at 95\% at ATLAS (blue) and CMS (violet), assuming 3 ab$^{-1}$.  Regions below solid lines and color shaded can yield successful baryogenesis with $H_{a,b}$ mass degeneracy corresponding to $m_a/m_b=0.95$ (magenta) and $m_a/m_b=0.9$ (turquoise).  Parameter space above the dotted lines and either gray or light color shaded are excluded by the flavor bounds  (Fig.~\ref{fig:flavorbounds}).}
	\label{fig:exc}
\end{figure*}     

Since the heavy Higgs states $H_a$ have substantial couplings to light quarks, they can be resonantly produced at hadron colliders.  Here, we will again assume the alignment limit for the Higgs sector.  Hence, $H_a$ do not decay into $H_1H_1$ or gauge boson pairs at leading order, and their dominant decay modes are into di-jet and $t\bar t$ finals states.  

For our HL-LHC projections, we assume 3 ab$^{-1}$ of data and use {\small $\sqrt{\tt{Luminosity}}$} scaling of existing ATLAS \cite{Aad:2019hjw} and CMS \cite{Sirunyan:2019vgj} bounds on the product of cross section, branching fraction, and acceptance.  To calculate the production cross section, the couplings of Eq.~(\ref{eq:flavorscheme}) are implemented in \texttt{MadGraph5\_aMC@NLO}~\cite{Alwall:2014hca} via \texttt{FeynRules}~\cite{Christensen:2008py,Alloul:2013bka}.  The acceptance is estimated using parton-level acceptance cuts in~\texttt{MadGraph5\_aMC@NLO}.

For a 100 TeV $pp$ collider, we adapt the $Z_B$ bound projections in Ref.~\cite{Golling:2016gvc}, by implementing the $Z_B$ model in~\texttt{MadGraph5\_aMC@NLO}.  We then map the gauge coupling versus mass limits to those for dijet cross sections.  All components of $H_a$ are assumed to have the same mass here.  To find the regions of parameter space that can lead to successful baryogenesis, we use the formalism developed in Ref.~\cite{Davoudiasl:2019lcg}  The asymmetry parameter is given by 
\begin{eqnarray}
\varepsilon = \frac{1}{8\pi}\frac{m_a^2}{m_b^2-m_a^2}
\frac{\sum_{f=u,d}N_{c,f}\text{Im}\left({\rm Tr}^{ba}_\nu {\rm Tr}^{ba*}_{f}\right)}{N_{c,f}{\rm Tr}^{aa}_f}\,,
\label{eq:eps}
\end{eqnarray}    
where ${\rm Tr}^{ba}_f={\rm Tr}[\lambda^{b\dagger}_{f}\lambda^a_f]$ and $N_c=3,1$ for quarks and leptons, respectively.  Assuming contributions from all quarks and using the washout condition in \eq{Ha-washout}, we find that approximately  
\beq
\varepsilon < 1.8\times10^{-9}\left(\frac{m_a}{10 {\rm TeV}}\right)^4\label{eq:epsnum}
\frac{1}{[(m_b/m_a)^2-1](\kappa_d^2+\kappa_s^2+\kappa_d^2+\xi^2)}\,.
\eeq
In the above, a physical phase $\theta_f$, with $|\sin \theta_f| \leq 1$, is assumed in the product of fermion couplings, so that $\eps\neq 0$.

For $H_a$ masses up to 30 TeV considered here, the $\Phi$ modulus mass $m_\Phi > 60$ TeV and Eq.~(\ref{eps-mPhi}) implies that $\varepsilon >  10^{-7}$; the shaded regions in Fig.~\ref{fig:exc} correspond to this baryogenesis requirement.  For a given Higgs mass and $\xi$, Eq.~(\ref{Ha-washout}) sets a maximum neutrino coupling, while $\varepsilon > 10^{-7}$ gives a minimum.  As we can see from Fig.~\ref{fig:exc}, the discovery reach for the Higgs Troika baryogenesis model adopted here is quite impressive at a 100 TeV $pp$ collider and can probe scalar masses $\ord{10~\rm TeV}$.  Also, our results show that flavor data can play a complementary role in searching for the signals of the scenario discussed here.  In particular, improved $D-\bar{D}$ mixing constraints can be quite sensitive for $\kappa_d=\kappa_s=\kappa_b$, $\xi=1$ and $m_a > 15-20$ TeV. 

The above estimates correspond to search for one Higgs doublet.  However, the Troika baryogenesis model requires at least two heavy Higgs doublets.  We end this summary with a few remarks about effect of $H_a$ widths and spectrum on the discovery reach.  If $H_2$ and $H_3$ are well-separated in mass, we can expect distinct resonances.  However, if the Higgs masses differ by more than the widths, but less than the detector jet energy resolution, we may expect a broad resonance.  This could enhance the cross section reach by a factor $\sim \sqrt{2}$.  If the Higgs widths are larger than the mass separation, we may expect an enhanced sensitivity up to a factor of $\sim 2$, due to coherence.

\section*{Acknowledgments}

The author would like to thank the organizers of the virtual Rencontres de Moriond 2021 meeting for providing the opportunity to present the above results, and Ian M. Lewis and Mathew Sullivan, for collaboration on the work summarized here.   H.D. and M.S. are supported by the United States Department of Energy under Grant Contract DE-SC0012704.  I.M.L. is supported in part by the United States Department of Energy grant number DE-SC0017988.   

%

\section*{References}

\bibliography{higgstroika-moriond2021-refs}

\begin{thebibliography}{10}

\bibitem{Davoudiasl:2019lcg}
Hooman Davoudiasl, Ian~M. Lewis, and Matthew Sullivan.
\newblock {Higgs Troika for Baryon Asymmetry}.
\newblock {\em Phys. Rev. D}, 101(5):055010, 2020.

\bibitem{Davoudiasl:2021syn}
Hooman Davoudiasl, Ian~M. Lewis, and Matthew Sullivan.
\newblock {Multi-TeV Signals of Baryogenesis in Higgs Troika Model}.
\newblock 2103.12089.

\bibitem{Fukugita:1986hr}
M.~Fukugita and T.~Yanagida.
\newblock {Baryogenesis Without Grand Unification}.
\newblock {\em Phys. Lett. B}, 174:45--47, 1986.

\bibitem{Sakharov:1967dj}
A.D. Sakharov.
\newblock {Violation of CP Invariance, C asymmetry, and baryon asymmetry of the
  universe}.
\newblock {\em Sov. Phys. Usp.}, 34(5):392--393, 1991.

\bibitem{Dick:1999je}
Karin Dick, Manfred Lindner, Michael Ratz, and David Wright.
\newblock {Leptogenesis with Dirac neutrinos}.
\newblock {\em Phys. Rev. Lett.}, 84:4039--4042, 2000.

\bibitem{Murayama:2002je}
Hitoshi Murayama and Aaron Pierce.
\newblock {Realistic Dirac leptogenesis}.
\newblock {\em Phys. Rev. Lett.}, 89:271601, 2002.

\bibitem{Davoudiasl:2011aa}
Hooman Davoudiasl and Ian Lewis.
\newblock {Technicolor Assisted Leptogenesis with an Ultra-Heavy Higgs
  Doublet}.
\newblock {\em Phys. Rev. D}, 86:015024, 2012.

\bibitem{Cheng:1987rs}
T.~P. Cheng and Marc Sher.
\newblock {Mass Matrix Ansatz and Flavor Nonconservation in Models with
  Multiple Higgs Doublets}.
\newblock {\em Phys. Rev. D}, 35:3484, 1987.

\bibitem{Grossman:1994jb}
Yuval Grossman.
\newblock {Phenomenology of models with more than two Higgs doublets}.
\newblock {\em Nucl. Phys. B}, 426:355--384, 1994.

\bibitem{Cree:2011uy}
Graham Cree and Heather~E. Logan.
\newblock {Yukawa alignment from natural flavor conservation}.
\newblock {\em Phys. Rev. D}, 84:055021, 2011.

\bibitem{Ivanov:2011ae}
Igor~P. Ivanov, Venus Keus, and Evgeny Vdovin.
\newblock {Abelian symmetries in multi-Higgs-doublet models}.
\newblock {\em J. Phys. A}, 45:215201, 2012.

\bibitem{Keus:2013hya}
Venus Keus, Stephen~F. King, and Stefano Moretti.
\newblock {Three-Higgs-doublet models: symmetries, potentials and Higgs boson
  masses}.
\newblock {\em JHEP}, 01:052, 2014.

\bibitem{Bento:2017eti}
Miguel~P. Bento, Howard~E. Haber, J.~C. Rom\~ao, and Jo\~ao~P. Silva.
\newblock {Multi-Higgs doublet models: physical parametrization, sum rules and
  unitarity bounds}.
\newblock {\em JHEP}, 11:095, 2017.

\bibitem{Alves:2020brq}
Jo\~ao~M. Alves, Francisco~J. Botella, Gustavo~C. Branco, and Miguel Nebot.
\newblock {Extending trinity to the scalar sector through discrete flavoured
  symmetries}.
\newblock {\em Eur. Phys. J. C}, 80(8):710, 2020.

\bibitem{Logan:2020mdz}
Heather~E. Logan, Stefano Moretti, Diana Rojas-Ciofalo, and Muyuan Song.
\newblock {CP violation from charged Higgs bosons in the three Higgs doublet
  model}.
\newblock 12 2020.

\bibitem{Egana-Ugrinovic:2018znw}
Daniel Egana-Ugrinovic, Samuel Homiller, and Patrick Meade.
\newblock {Aligned and Spontaneous Flavor Violation}.
\newblock {\em Phys. Rev. Lett.}, 123(3):031802, 2019.

\bibitem{Egana-Ugrinovic:2019dqu}
Daniel Egana-Ugrinovic, Samuel Homiller, and Patrick~Roddy Meade.
\newblock {Higgs bosons with large couplings to light quarks}.
\newblock {\em Phys. Rev. D}, 100(11):115041, 2019.

\bibitem{Andreev:2018ayy}
V.~Andreev et~al.
\newblock {Improved limit on the electric dipole moment of the electron}.
\newblock {\em Nature}, 562(7727):355--360, 2018.

\bibitem{Aad:2019hjw}
Georges Aad et~al.
\newblock {Search for new resonances in mass distributions of jet pairs using
  139 fb$^{-1}$ of $pp$ collisions at $\sqrt{s}=13$ TeV with the ATLAS
  detector}.
\newblock {\em JHEP}, 03:145, 2020.

\bibitem{Sirunyan:2019vgj}
Albert~M Sirunyan et~al.
\newblock {Search for high mass dijet resonances with a new background
  prediction method in proton-proton collisions at $\sqrt{s} =$ 13 TeV}.
\newblock {\em JHEP}, 05:033, 2020.

\bibitem{Alwall:2014hca}
J.~Alwall, R.~Frederix, S.~Frixione, V.~Hirschi, F.~Maltoni, O.~Mattelaer,
  H.~S. Shao, T.~Stelzer, P.~Torrielli, and M.~Zaro.
\newblock {The automated computation of tree-level and next-to-leading order
  differential cross sections, and their matching to parton shower
  simulations}.
\newblock {\em JHEP}, 07:079, 2014.

\bibitem{Christensen:2008py}
Neil~D. Christensen and Claude Duhr.
\newblock {FeynRules - Feynman rules made easy}.
\newblock {\em Comput. Phys. Commun.}, 180:1614--1641, 2009.

\bibitem{Alloul:2013bka}
Adam Alloul, Neil~D. Christensen, C\'eline Degrande, Claude Duhr, and Benjamin
  Fuks.
\newblock {FeynRules 2.0 - A complete toolbox for tree-level phenomenology}.
\newblock {\em Comput. Phys. Commun.}, 185:2250--2300, 2014.

\bibitem{Golling:2016gvc}
T.~Golling et~al.
\newblock {Physics at a 100 TeV pp collider: beyond the Standard Model
  phenomena}.
\newblock {\em CERN Yellow Rep.}, (3):441--634, 2017.

\end{thebibliography}

%
%
%
%

\end{document}